\documentclass{jpsj2}
\title{Anomalous electric conductions in KSbO$_{3}$-type metallic rhenium oxides}

\author{Hirotake \textsc{Suzuki}, Hiromi \textsc{Ozawa} and Hirohiko \textsc{Sato}\thanks{E-mail address:
hirohiko@phys.chuo-u.ac.jp}
}

\inst{Department of Physics, Chuo University, 1-13-27 Kasuga, Bunkyo-ku,
Tokyo 112-8551, Japan
 \\
}

\abst{
Single crystals of KSbO$_{3}$-type rhenium oxides, La$_{4}$Re$_{6}$O$_{19}$, Pb$_{6}$Re$_{6}$O$_{19}$, Sr$_{2}$Re$_{3}$O$_{9}$ and Bi$_{3}$Re$_{3}$O$_{11}$, were synthesized by a hydrothermal method. Their crystal structures can be regarded as a network of three-dimensional orthogonal-dimer lattice of edge-shared ReO$_{6}$ octahedra. All of them exhibit small magnitude of Pauli paramagnetism, indicating metallic electronic states without strong electron correlations. The resistivity of these rhenates, except Bi$_{3}$Re$_{3}$O$_{11}$, have a temperature dependence of $\rho(T)=\rho_{0}+AT^{n}$ $(n \approx 1.6)$ in a wide temperature range between 5 K and 300 K, which is extraordinary for three-dimensional metals without strong electron correlations. The resistivity of Bi$_{3}$Re$_{3}$O$_{11}$ shows an anomaly around at 50 K, where the magnetic susceptibility also detects a deviation from ordinary Pauli paramagnetism. 
}

\kword{transition metal oxide, rhenium oxide, magnetism, orthogonal dimer lattice, crystal structure, electoric conduction
}

\begin{document}
\maketitle

\section{Introduction}
There are many rhenium oxides (rhenates) exhibiting metallic electric conductions. The binary rhenate, ReO$_{3}$, composed of isotropic three-dimensional network of vertex-shard ReO$_{6}$ octahedra, is well known for good electric conductivity as high as that of Cu metal.\cite{ferretti65,tanaka76} In this rhenate, each Re$^{6+}$ ion donates an electron to the wide band composed of the large $5d$ orbitals.\cite{mattheiss69} More than 30 years after, rhenates became one of the topical oxides again because of the discovery of superconductivity in Cd$_{2}$Re$_{2}$O$_{7}$.\cite{sakai01,hanawa01} Although its critical temperature is as low as 1 K and the pairing mechanism is not unconventional one, the discovery is important because Cd$_{2}$Re$_{2}$O$_{7}$ is the first superconductor with a pyrochlore structure. Apart from the superconductivity, Cd$_{2}$Re$_{2}$O$_{7}$ is also an interesting materials for its successive phase transitions related to the geometric frustration inherent in pyrochlores.

There are a series of rhenates with another type of structure, so called KSbO$_{3}$-type. In this structure, the basic unit is an edge-shared ``dimer'' of two ReO$_{6}$ octahedra. These dimers are further connected to each other, sharing the corner oxygen atoms. One of the most interesting features in this structure is that the direction of the elongation of the adjacent dimers are perpendicular to each other. Therefore, the Re sites form a three-dimensional version\cite{chen02} of orthogonal dimer lattice which has been theoretically investigated in two-dimensional case by Shastry and Sutherland.\cite{shastry81} In the two-dimensional case, Kageyama \emph{et al}. synthesized a modle compound SrCu$_{2}$(BO$_{3}$)$_{2}$ and discovered a spin-singlet ground state without a long-range magnetic order. In high magnetic field, characteristic plataus appear in the magnetization curve.\cite{kageyama99,miyahara03} Also in a three-dimensional orthogonal-dimer system, similar interesting magnetism is expected if there is a localized spin on each site. Unfortunately, all of KSbO$_{3}$-type transition-metal oxides whose physical properties have been investigated are metals without localized spins. Nevertheless, exotic electronic states have been reported in several KSbO$_{3}$-type oxides, although the relationship with the singular structure has not necessarily been clarified. The conductivity and the magnetism of La$_{4}$Ru$_{6}$O$_{19}$ exhibit anomalous behaviors suggesting non-Fermi liquid states.\cite{khalifah01,khalifah01b} Similar anomalies were also reported in Bi$_{3}$Ru$_{3}$O$_{11}$ and Bi$_{3}$Os$_{3}$O$_{11}$.\cite{fujita03,tsuchida04} An extraordinary temperature dependence of magnetism, which remind us a spin-gap system, has been also discovered by our group in (Ba$_{1-x}$Sr$_{x}$)$_{2}$Ru$_{3}$O$_{9}$.\cite{sato04}

In the case of rhenates, Sr$_{x}$ReO$_{3}$ ($0.4 \le x \le 0.5$),\cite{baud79} Pb$_{6}$Re$_{6}$O$_{19}$,\cite{abakumov98} La$_{4}$Re$_{6}$O$_{19}$,\cite{longo68,morrow68} Ln$_{4}$Re$_{6-x}$O$_{19}$\cite{bramnik98,jeitschko99} and Bi$_{3}$Re$_{3}$O$_{11}$,\cite{cheetham81} have been known to have KSbO$_{3}$ structure, but their physical properties have not been investigated, except for the recent studies on powder samples of La$_{4}$Re$_{6}$O$_{19}$, Pr$_{4}$Re$_{6}$O$_{19}$ and Nd$_{4}$Re$_{6}$O$_{19}$.\cite{jeitschko99,sasaki06} In the present study, we synthesized single crystals of La$_{4}$Re$_{6}$O$_{19}$, Pb$_{6}$Re$_{6}$O$_{19}$, Sr$_{2}$Re$_{3}$O$_{9}$ and Bi$_{3}$Re$_{3}$O$_{11}$ by a hydrothermal method. Refinements of the crystal structure were performed on Sr$_{2}$Re$_{3}$O$_{9}$ and Bi$_{3}$Re$_{3}$O$_{11}$. We also measured the resistivities and the magnetic susceptibilities on all of the KSbO$_{3}$-type rhenates we obtained in this study. It was revealed that all of them exhibit metallic behaviors without strong electron correlations. However, their resistivities have anomalous temperature dependences.

\section{Experimental}
La$_{4}$Re$_{6}$O$_{19}$, Pb$_{6}$Re$_{6}$O$_{19}$, Sr$_{2}$Re$_{3}$O$_{9}$ and Bi$_{3}$Re$_{3}$O$_{11}$ were synthesized by a hydrothermal method. A mixture of the starting materials was sealed in a silver capsule with water. Then the capsule was put in a reactor filled with high-pressure water. The reactor was heated in a furnace, maintaining 150 MPa of hydrostatic pressure for three days. The obtained crystals were washed with hot water. The chemical compositions were analyzed on single crystals using an energy dispersive X-ray spectrometer (EDS) (Oxford, Inca Energy 500) installed on a scanning electron microscope. The crystal structure was analyzed using an imaging-plate type X-ray diffractometer (Rigaku, RAPID R-Axis).

Magnetic susceptibilities were measured on an ensamble of single crystals using a superconducting-quantum-intereference-device (SQUID) magnetometer (Quantum Design, MPMS-XL). The samples were wrapped with a piece of aluminum foil and it was fixed in a straw. The magnetic susceptibility of the aluminum foil, whose temperature dependence was carefully examined in the independent measurement, was subtracted from the data. The resistivities were measured on a single crystal by DC four-terminal method. Gold wires were attached on a single crystal with carbon paste. Only in the case of Pb$_{6}$Re$_{6}$O$_{19}$, we attached the wires on polycrystals, composed of about ten single crystals, because of the very small size of the sample. A closed-cycle helium refrigerator or the SQUID system was used for the resistivity measurements. We also measured AC magnetic susceptibilities down to 50 mK using a demagnetization cryostat (CMR, m-Frige) just for searching superconducting signals.

\section{Results and discussion}
\subsection{Characterization and crystal structures}
\begin{figure}[tb]
\begin{center}
\includegraphics[width=1.0\linewidth]{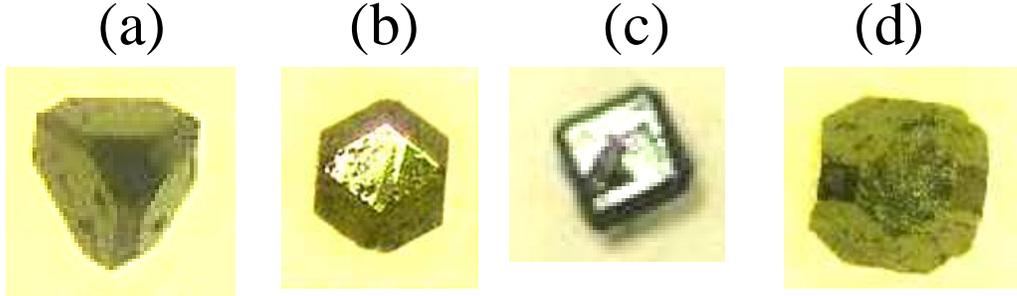}
\end{center}
\caption{Microscope photographs of (a) La$_{4}$Re$_{6}$O$_{19}$, (b) Pb$_{6}$Re$_{6}$O$_{19}$, (c) Sr$_{2}$Re$_{3}$O$_{9}$ and (d) Bi$_{3}$Re$_{3}$O$_{11}$. The typical size of the crystals were $0.2\times 0.2 \times 0.2$ mm$^{2}$.}
\label{fig1}
\end{figure}

Table~\ref{table1} summarizes the conditions of the hydrothermal syntheses and the results. In any cases, we were able to obtain single crystals, which were black-colored truncated cubes as shown in Fig.~\ref{fig1}. The typical size of the crystals were $0.2\times 0.2 \times 0.2$ mm$^{3}$. As listed in the table, analyses of the chemical composition of non-oxygen elements were performed on single crystals using EDS. X-ray diffraction measurements on a single crystal revealed cubic symmetry for all of the materials. The lattice constants, $a$, are also listed in the Table~\ref{table1}.

\begin{table}
\caption{Synthetic conditions and results of KSbO$_{3}$-type rhenates. The mixture of starting materials (indicated in molar ratios) are sealed with water and reacted in a hydrothermal condition. EDS analyses were performed on the obtained single crystals. The lattice constants of the material 3 and 4 were determined from our single-crystal X-ray diffraction measurements.}
\label{table1}
\begin{tabular}{ccccc}
\hline
No. & Starting materials & $T$ ($^{\circ}$C) & Atomic ratio (EDS) & $a$ (\AA) \\
\cline{1-5}
1 & ReO$_{2}$:La$_{2}$O$_{3} = 1:1$ & 650 & La:Re $ = 2:3$ & 9.038\cite{longo68} \\
2 & ReO$_{2}$:PbO $ = 1:3$ & 500 &  Pb:Re $ = 1:1$ & 9.318\cite{abakumov98} \\
3 & ReO$_{2}$:SrO$_{2} = 1:1$ & 580 & Sr:Re $ = 2:3$ & 9.226 \\
4 & ReO$_{2}$:Bi$_{2}$O$_{3} = 1:1$ & 600 & Bi:Re $ =1:1$ & 9.367 \\
\hline
\end{tabular}
\end{table}

From the compositions and the lattice constants, the material 1 and 2 were easily identified as La$_{4}$Re$_{6}$O$_{19}$\cite{longo68,morrow68} and Pb$_{6}$Re$_{6}$O$_{19}$\cite{abakumov98}, respectively. The material 3 has a slightly larger lattice constant than that of reported Sr$_{x}$ReO$_{3}$ ($a=9.192$ \AA)\cite{baud79}. In addition, the Sr content is significantly larger in our material than that in Sr$_{x}$ReO$_{3}$ ($0.4 \le x \le 0.5$).\cite{baud79} Therefore, we name our material Sr$_{2}$Re$_{3}$O$_{9}$ here and distinguish it from theirs. The composition and the lattice constant of material 4 coincide with those of Bi$_{3}$Re$_{3}$O$_{11}$\cite{cheetham81} whose detailed structure has not been investigated at all. Therefore, we carried out refinements of the crystal structures on the Sr$_{2}$Re$_{3}$O$_{9}$ and Bi$_{3}$Re$_{3}$O$_{11}$ using the single crystal X-ray diffraction data. The results are summerized in Table~\ref{table2} and Table~\ref{table3}. In the structural refinement of Bi$_{3}$Re$_{3}$O$_{11}$, we fixed the temperature factors of Re and O, otherwise they became negative value. This problem was caused by the strong absorption coefficient of X-ray by heavy Bi and Re elements, and exact correction for this effect was very difficult.

\begin{table}[tb]
\caption{Fractional atomic coordinates and equivalent isotropic displacement parameters 
 (\AA$^{2}$) for Sr$_{2}$Re$_{3}$O$_{9}$. The lattice symmetry and the space group 
are \textit{cubic} and $Im\bar{3}$ (\#204), respectively. The lattice parameters are  $a = 9.226(1)$ \AA{}, 
$V = 785.4(2)$ \AA{}$^3$ and $Z=4$. The final reliability factor is $R(F) = 10.4 \%$.($I>3\sigma$)}
\label{table2}
\begin{tabular}{ccccccc}
\hline
Atom & Position & $x$ & $y$ & $z$ & $B_{eq}$ & occ.\\
\cline{1-7}
Re(1) & 12$e$ & 0.36843(4) & 0.0000   &  0.5000 &    0.240(5) &  1/4\\
Sr(1) & 16$f$ & 0.3568(3) & 0.3568(3) &  0.6433(3) & 0.56(5) &  0.071\\
Sr(2) & 16$f$ & 0.3047(3) & 0.3047(3) &  0.6953(3) & 0.71(5) &  0.073\\
Sr(3) & 8$c$  & 0.2500 &    0.2500 &     0.7500 & 0.5(1) &  0.023\\
O(1)  & 24$g$ & 0.2122(7) & 0.8500(7) & 0.5000 & 0.68(8) &  1/2\\
O(2)  & 12$d$ & 0.5000 & 0.1710(9) &     0.5000 & 0.5(1) &  1/4\\
\hline
\end{tabular}
\end{table}

\begin{table}[tb]
\caption{Fractional atomic coordinates and equivalent isotropic displacement parameters 
 (\AA$^{2}$) for Bi$_{3}$Re$_{3}$O$_{11}$. The lattice symmetry and the space group 
are \textit{cubic} and $Pn\bar{3}$ (\#201), respectively. The lattice parameters are  $a = 9.367(2)$ \AA{}, 
$V = 821.9(3)$ \AA{}$^3$ and $Z=4$. The final reliability factor is $R(F) = 4.4 \%$.($I>3\sigma$)}
\label{table3}
\begin{tabular}{ccccccc}
\hline
Atom & Position & $x$ & $y$ & $z$ & $B_{eq}$ & occ.\\
\cline{1-7}
Bi(1) & 4$b$ & 0 & 0   &  0 &    0.41(2) &  1/6\\
Bi(2) & 8$e$ & 0.38243(9) & 0.38243(9) & 0.38243(9) & 0.29(1) &  1/3\\
Re(1) & 12$g$  & .3816(1) &    0.75 &     0.25 & 0.2 &  1/2\\
O(1)  & 12$f$ & 0.588(3) & 0.25 & 0.25 & 1.0 &  1/2\\
O(2)  & 8$e$ & 0.145(2) & 0.145(2) & 0.145(2) & 1.0 &  1/3\\
O(3)  & 24$h$ & 0.601(2) & 0.245(2) & 0.532(2) & 1.0 &  1\\
\hline
\end{tabular}
\end{table}

\begin{figure}[tb]
\begin{center}
\includegraphics[width=1.0\linewidth]{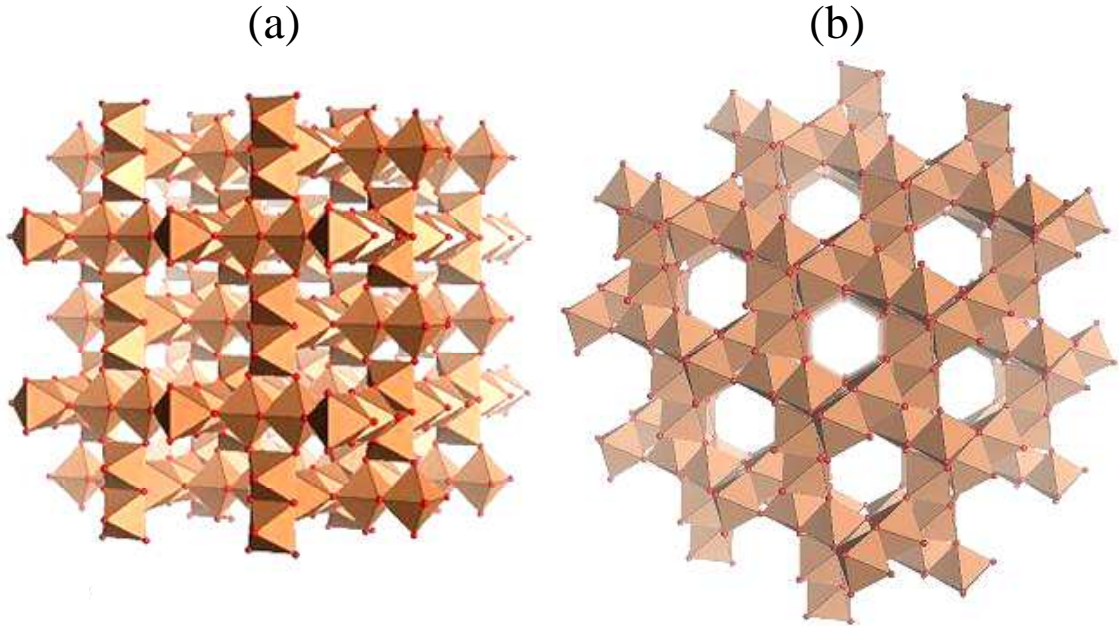}
\end{center}
\caption{Re-O framework structure in KSbO$_{3}$-type rhenates. (a) A view from almost (100) direction. The ReO$_{6}$ octahedra form a three-dimensional orthogonal dimer lattice. (b) A view from (111) direction. Tunnel-like necks are seen.}
\label{fig2}
\end{figure}

\begin{figure}[tb]
\begin{center}
\includegraphics[width=1.0\linewidth]{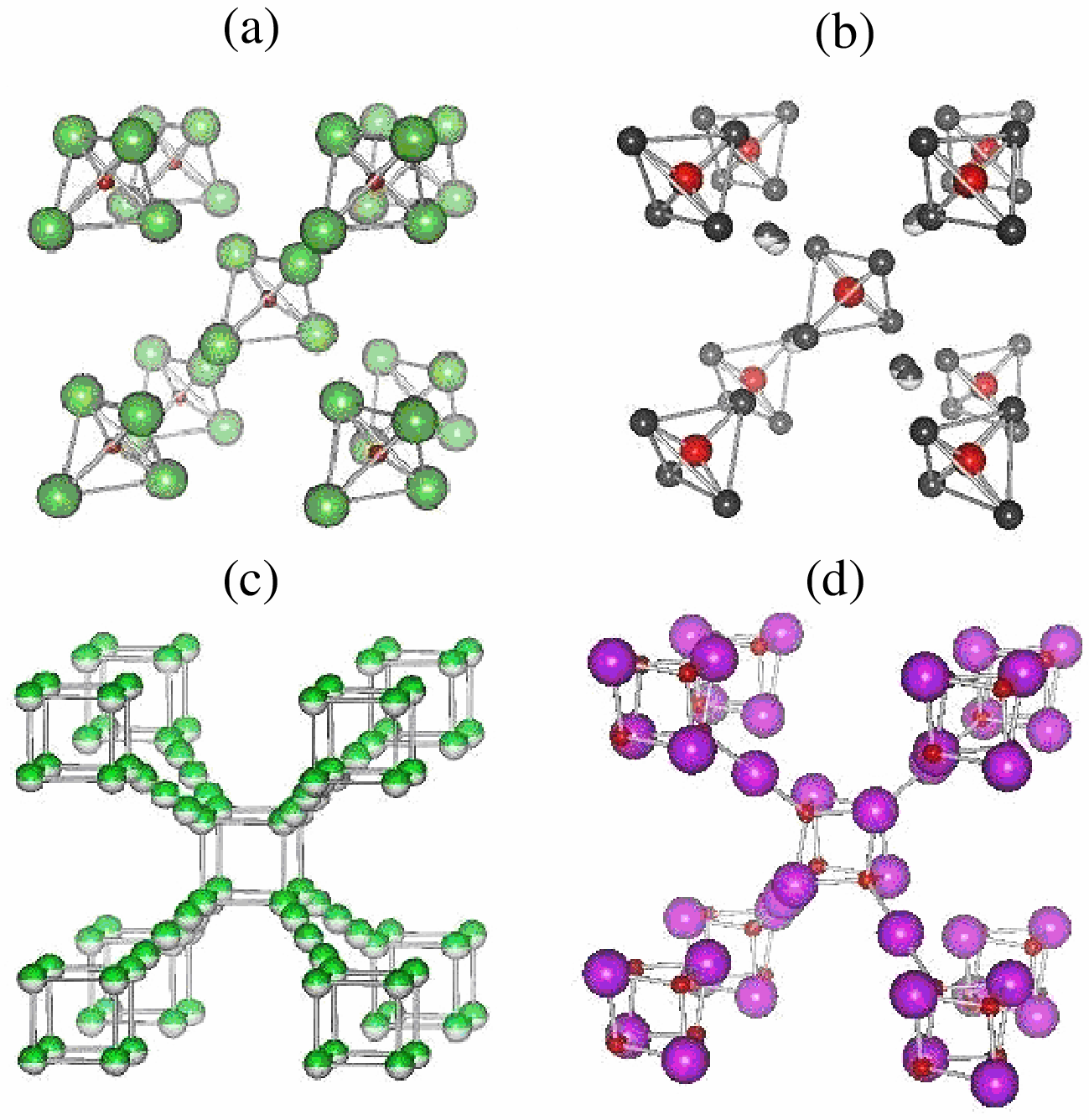}
\end{center}
\caption{Comparison of cation arrengiments in (a) La$_{4}$Re$_{6}$O$_{19}$, (b) Pb$_{6}$Re$_{6}$O$_{19}$, (c) Sr$_{2}$Re$_{3}$O$_{9}$ and (d) Bi$_{3}$Re$_{3}$O$_{11}$. The partially occupied sites are indicated with two-tone-colored spheres. In La$_{4}$Re$_{6}$O$_{19}$, the La$_{4}$O tetrahedra occupy the pores. Pb$_{6}$Re$_{6}$O$_{19}$ has additional cations in the necks. In Sr$_{2}$Re$_{3}$O$_{9}$, Sr ions are randomly distributed on many sites in the pores and the necks. Bi$_{3}$Re$_{3}$O$_{11}$ has the Bi$_{3}$O$_{3}$ distorted cubes in the pores and Bi ions in the necks.}
\label{fig3}
\end{figure}

La$_{4}$Re$_{6}$O$_{19}$, Pb$_{6}$Re$_{6}$O$_{19}$, Sr$_{2}$Re$_{3}$O$_{9}$ and Bi$_{3}$Re$_{3}$O$_{11}$ have orthogonal dimer lattice of Re network as shown in Fig.~\ref{fig2}(a). Two ReO$_{6}$ octahedra form a Re$_{2}$O$_{10}$ dimer sharing their edges. These dimers further form a three-dimensional netowork, sharing the corner oxygens. The directions of the long axis of adjacent dimers are perpendicular to each other. This Re-O framework is so porous that large spherical spaces surrounded by the cage of Re-O framework are left. We call them pores. The pores are linked by tunnels, elongated along the [111] and equivalent directions (Fig.~\ref{fig2}(b)), each other. We name them necks. The guest ions, such as the cations and the extra oxygen atoms, are located at the pores and/or necks. There are a variety of the arrangements of the guest ions as shown in Fig.~\ref{fig3}. In La$_{4}$Re$_{6}$O$_{19}$, the La ions form a La$_{4}$O tetrahedral cluster in a pore. Pb$_{6}$Re$_{6}$O$_{19}$ has a similar arrangement of guest ions as that in La$_{4}$Re$_{6}$O$_{19}$, but there are extra Pb ions in the necks with 1/2 occupancy. Our structural analysis revealed that Sr$_{2}$Re$_{3}$O$_{9}$ has more disordered guest ion arrangement. The Sr ions randomly occupy many sites spread in the pores and the necks. In Bi$_{3}$Re$_{3}$O$_{11}$, there is no disorder in the cation arrangement. Bi and O form distorted cubes in the pores and extra Bi ions are located in the necks. The intradimer and interdimer Re-Re distances as summerized in Table~\ref{table4}.

\begin{table}
\caption{Interatomic distances (\AA) in KSbO$_{3}$-type rhenates.}
\label{table4}
\begin{tabular}{ccccc}
\hline
bonding & La$_{4}$Re$_{6}$O$_{19}$ & Pb$_{6}$Re$_{6}$O$_{19}$ & Sr$_{2}$Re$_{3}$O$_{9}$ & Bi$_{3}$Re$_{3}$O$_{11}$ \\
\cline{1-5}
Re-Re (intradimer) & 2.414  & 2.450  & 2.428  & 2.465  \\
Re-Re (interdimer) & 3.525 & 3.646  & 3.609  & 3.664  \\
\hline
\end{tabular}
\end{table}

In the case of ruthenates, Khalifah, \emph{et al}. points out that La$_{4}$Ru$_{6}$O$_{19}$ has extremely short intradimer Ru-Ru distance, 2.488 \AA. Therefore, there is a direct Ru-Ru bonding and, consequently, the band structure should be understood based on the molecular orbitals of edge-shared Ru$_{2}$O$_{10}$ cluster.\cite{khalifah01b} In our rhenates, the intradimer Re-Re distances are very short as listed in Table~\ref{table4}. Because the 5$d$ orbitals of Re extended wider than 4$d$ orbitals of Ru, formation of Re-Re direct bonding is reasonablily expected. This idea has been supported by the FAPW band calculation.\cite{sasaki06}

\subsection{Magnetism}

\begin{figure}[tb]
\begin{center}
\includegraphics[width=1.0\linewidth]{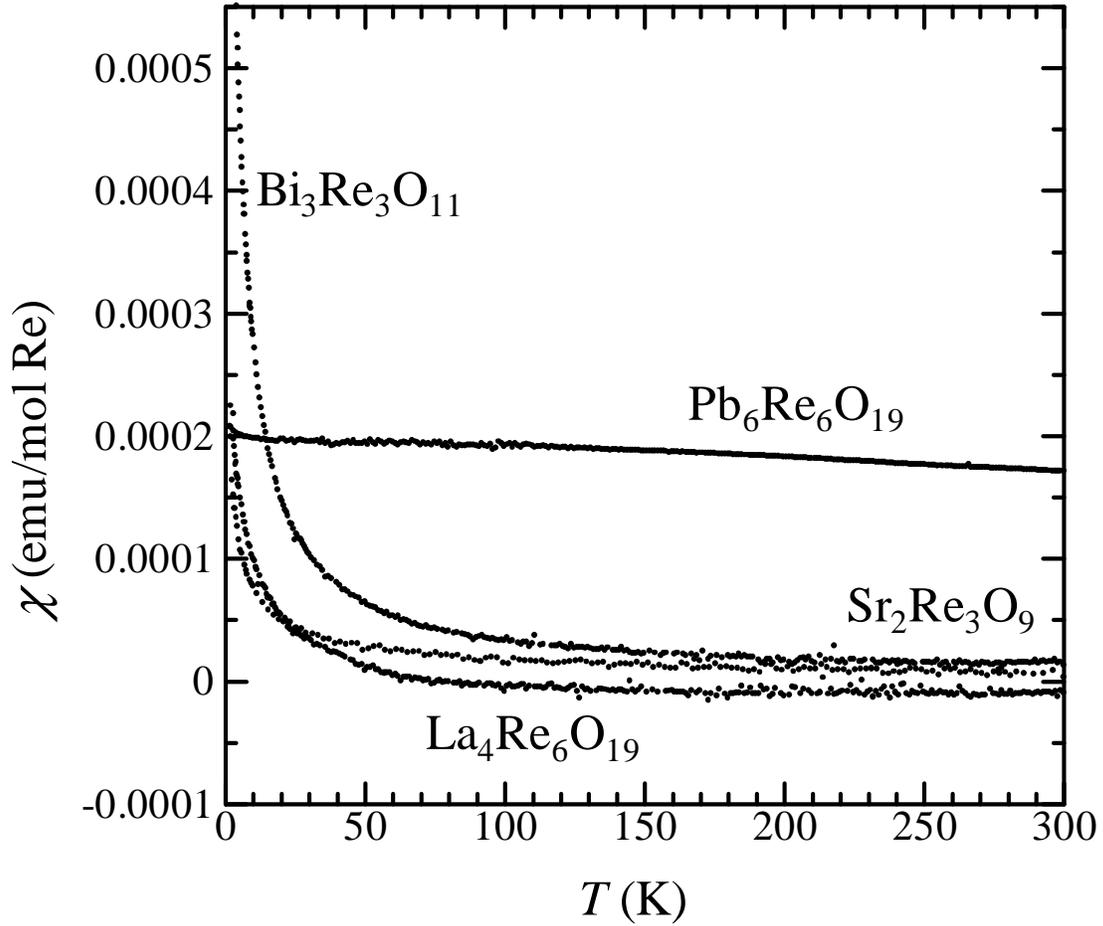}
\end{center}
\caption{Raw data of the magnetic susceptibilities of KSbO$_{3}$-type rhenates. Except for Pb$_{6}$Re$_{6}$O$_{19}$, Curie-Weiss behaviors appear at low temperature. The applied magnetic field was 1 T.}
\label{fig4}
\end{figure}

Figure~\ref{fig4} shows the raw data of the magnetic susceptibilities of the KSbO$_{3}$-type rhenates without any corrections. Because these rhenates are composed of heavy elements, the subtraction of closed-shell diamagnetism is essential. The calculated values of the closed-shell diamagnetism are $-7.9 \times 10^{-5}$, $-9.4 \times 10^{-5}$, $-7.4 \times 10^{-5}$ and $-9.4 \times 10^{-5}$ emu/mol~Re for La$_{4}$Ru$_{6}$O$_{19}$, Pb$_{6}$Re$_{6}$O$_{19}$, Sr$_{2}$Re$_{3}$O$_{9}$ and Bi$_{3}$Re$_{3}$O$_{11}$, respectively. Except for Pb$_{6}$Re$_{6}$O$_{19}$, the magnetic susceptibilities are strongly affected by Curie-Weiss contributions at low temperature. We estimated the Curie-Weiss contribution by least-square fittings below 20 K. The obtained Curie constants are $1.77 \times 10^{-3}$, $5.81 \times 10^{-4}$ and $4.70 \times 10^{-3}$ emu$\cdot$K/mol~Re for La$_{4}$Re$_{6}$O$_{19}$, Sr$_{2}$Re$_{3}$O$_{9}$ and Bi$_{3}$Re$_{3}$O$_{11}$, respectively. The Weiss temperatures are -5.1 K, -1.5 K and -3.3 K, respectively. The Curie constants are far smaller than what is expected if the $5d$ electrons of Re atoms are localized. Assuming that $S=1/2$ localized spins cause the Curie-Weiss behavior, the spin concentration is only 0.47 \%, 0.15\% and 1.3\% per Re site for La$_{4}$Re$_{6}$O$_{19}$, Sr$_{2}$Re$_{3}$O$_{9}$ and Bi$_{3}$Re$_{3}$O$_{11}$, respectively. Therefore, we can conclude that the $5d$ electrons are delocalized and that Curie-Weiss behaviors are not intrinsic in these materials but come from magnetic impurities or defects.

\begin{figure}[tb]
\begin{center}
\includegraphics[width=1.0\linewidth]{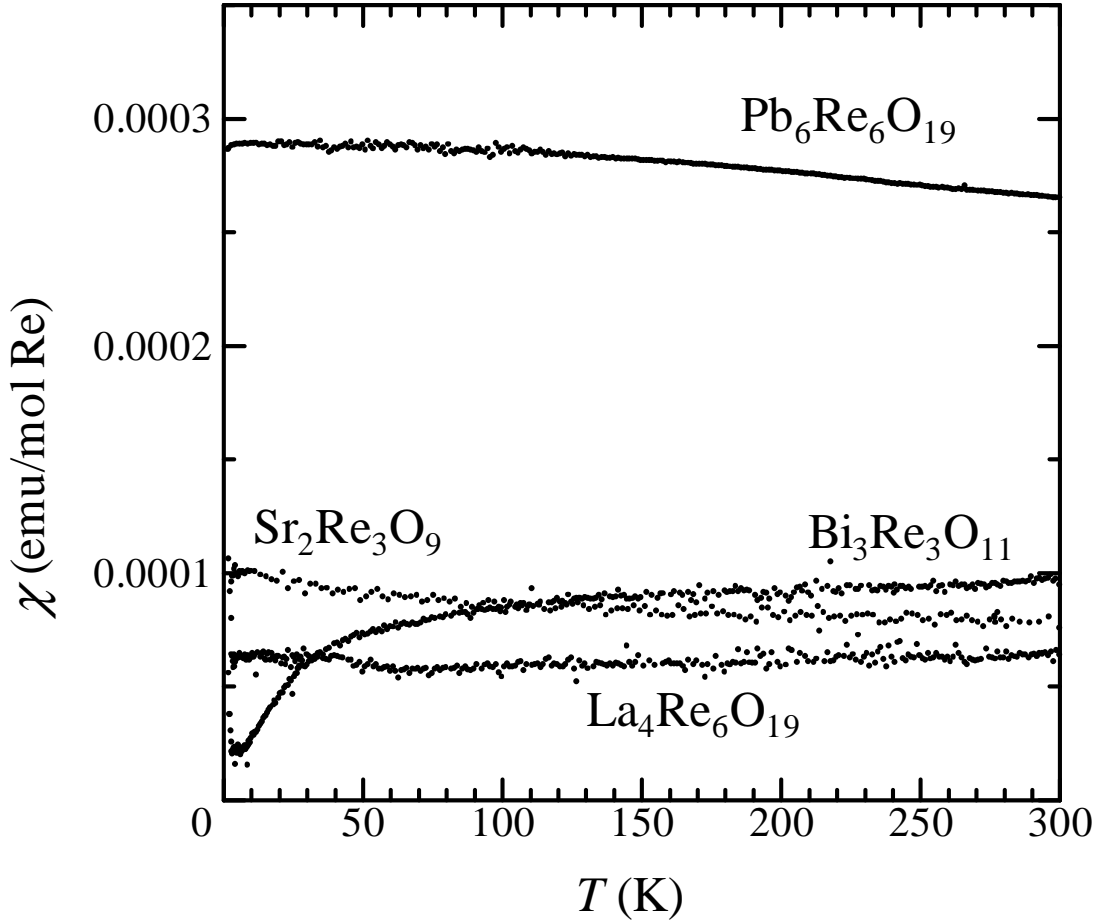}
\end{center}
\caption{Magnetic suscetibilities of KSbO$_{3}$-type rhenates after the subtraction of closed-shell diamagnetism ($-7.9 \times 10^{-5}$, $-9.4 \times 10^{-5}$, $-7.4 \times 10^{-5}$ and $-9.4 \times 10^{-5}$ emu/mol~Re for La$_{4}$Ru$_{6}$O$_{19}$, Pb$_{6}$Re$_{6}$O$_{19}$, Sr$_{2}$Re$_{3}$O$_{9}$ and Bi$_{3}$Re$_{3}$O$_{11}$, respectively) and Curie-Weiss contributions (except for Pb$_{6}$Re$_{6}$O$_{19}$).}
\label{fig5}
\end{figure}

We subtracted the closed-shell diamagnetism and the Curie-Weiss contributions (except for Pb$_{6}$Re$_{6}$O$_{19}$) from the raw data. The results of this correction are shown in Fig.~\ref{fig5}. For Sr$_{2}$Re$_{3}$O$_{9}$ and La$_{4}$Re$_{6}$O$_{19}$, positive, weakly temperature-dependent paramagnetism remains. The magnitude of the susceptibilities is typical one for Pauli paramagnetism of ordinary metals. The magnetism of Pb$_{6}$Re$_{6}$O$_{19}$ can be also interpreted as Pauli paramagnetism, although it is slightly temperature dependent. Let us compare the value with that of other KSbO$_{3}$-type $5d$ transition metal oxides. The magnetic susceptibility of Bi$_{3}$Os$_{3}$O$_{11}$\cite{fujita03} and Ba$_{2}$Ir$_{3}$O$_{9}$\cite{kawamura04} are $2.5 \times 10^{-4}$ emu / mol Os and $3.8 \times 10^{-4}$ emu / mol Ir, respectively. These are slightly larger than those in our rhenates. The larger value of Pauli paramagnetism in Pb$_{6}$Re$_{6}$O$_{19}$ suggests that the Pb bands contribute to the density-of-states (DOS) at $E_{F}$. Sasaki \emph{et al.} have reported the magnetic susceptibility of La$_{4}$Re$_{6}$O$_{19}$ powders, which shows almost constant paramagnetism as large as $4.5 \times 10^{-3}$ emu per one formula unit of La$_{4}$Re$_{6}$O$_{19}$. This value $7.5 \times 10^{-4}$ emu/mol~Re, divided by six for comparison with our data, seems extraordinarily larger than those of our La$_{4}$Re$_{6}$O$_{19}$ single crystals and other KSbO$_{3}$-type $5d$ oxides.

 In the case of Bi$_{3}$Re$_{3}$O$_{11}$, we did not succeeded reproducing the magnetic susceptibility with a simple summation of a Pauli paramagnetism and a Curie-Weiss function. The subtraction of the best-fitted Curie-Weiss contribution, which reproduces the rise of the susceptibility at low temperature very well, leaves anomalous magnetism; the susceptibility decreases with decreasing temperature below about 50 K.

We measured AC magnetic susceptibilities between 50 mK and 4 K on all the rhenates in this study. However, both of the real part and the imaginary part of the AC susceptibilities were almost independent of temperature. Therefore, we conclude that these rhenates do not undergo a superconduting transition down to 50 mK.

\subsection{Resistivities}

\begin{figure}[tb]
\begin{center}
\includegraphics[width=1.0\linewidth]{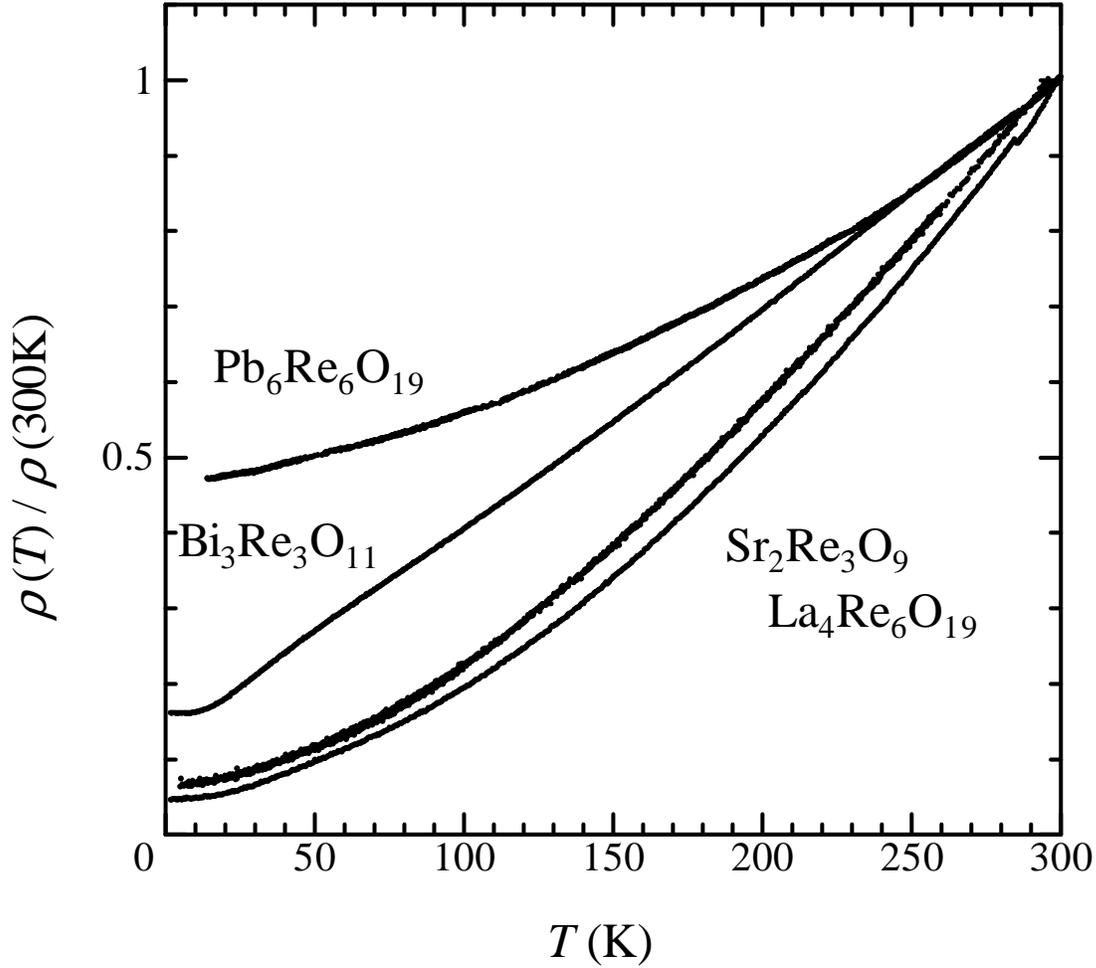}
\end{center}
\caption{Temperature dependences of the resistivity of KSbO$_{3}$-type rhenates. Single crystal samples were used for La$_{4}$Re$_{6}$O$_{19}$, Sr$_{2}$Re$_{3}$O$_{9}$ and Bi$_{3}$Re$_{3}$O$_{11}$. In case of Pb$_{6}$Re$_{6}$O$_{19}$, we used a polycrystalline sample composed of almost ten single crystals. Bi$_{3}$Re$_{3}$O$_{11}$ exhibits a round bend at 50 K.}
\label{fig6}
\end{figure}

\begin{figure}[tb]
\begin{center}
\includegraphics[width=1.0\linewidth]{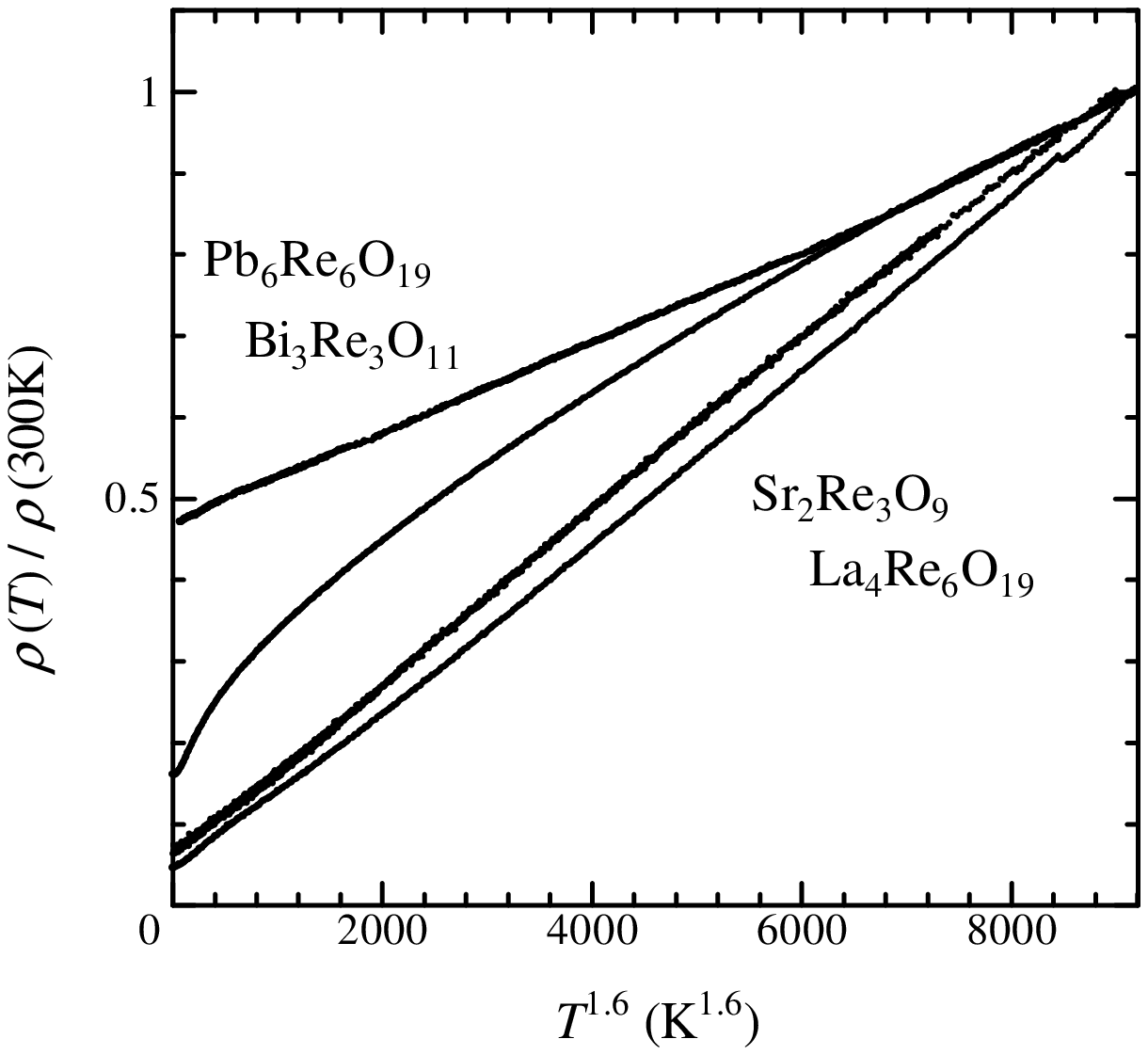}
\end{center}
\caption{The resistivities plotted against $T^{1.6}$. La$_{4}$Re$_{6}$O$_{19}$, Pb$_{6}$Re$_{6}$O$_{19}$ and Sr$_{2}$Re$_{3}$O$_{9}$ exhibit almost streight lines.}
\label{fig7}
\end{figure}

The resistivity of Pb$_{6}$Re$_{6}$O$_{19}$, La$_{4}$Re$_{6}$O$_{19}$, Sr$_{2}$Re$_{3}$O$_{9}$ and Bi$_{3}$Re$_{3}$O$_{11}$ at 300 K are as small as $4.6 \times 10^{-4}$, $2.5 \times 10^{-4}$, $8.1 \times 10^{-5}$ and $7.2 \times 10^{-4}$ $\Omega$ cm, respectively. As shown in Fig.~\ref{fig6}, the temperature dependence has positive slope in the whole temperature range, indicating metallic electronic states. The residual resitance ratio is smaller in Pb$_{6}$Re$_{6}$O$_{19}$ than in the other rhenates. This is because we used a polycrystal sample for Pb$_{6}$Re$_{6}$O$_{19}$, so that the contact resistance on the grain boundaries are not negligible. If the Fermi surface is spheric and the carriers are scattered by acoustic phonons, the temperature-dependent component of resistivities should be proportional to $T$ in high temperature region and to $T^{5}$ at low temperature, according to conventional Bloch-Gr\"{u}neisen model. However, the $\rho$ vs $T^{5}$ plots do not give straight lines at all. Instead, $\rho$ vs $T^{n}$ ($n = 1.6$) plots give more straight lines in a wide temperature range between 5 and 300 K, except for Bi$_{3}$Re$_{3}$O$_{11}$ as shown in Fig.~\ref{fig7}. The resistivity of the pelletized powder of La$_{4}$Re$_{6}$O$_{19}$ has been reported by Sasaki \emph{et al}.\cite{sasaki06} Except for larger residual resistance, similar anomalous temperature dependence, $\rho = \rho_{0}+AT^{1.39}$, appears also in the powder sample.  

The $T^{n}$ ($n \approx 2$) dependence of resistivity reminds us those in heavy Fermion systems. However, it is questionable to regard our rhenates as strong correlated electron systems, because the values of the Pauli paramagnetic susceptibilities are small. Furthermore, the band calculation and specific heat measurement on La$_{4}$Re$_{6}$O$_{19}$ indicate that the effective mass of the electrons is not enhanced.\cite{sasaki06} It should be noted that the $T^{2}$ dependence of resistivity has been recently reported in Ag$_{5}$Pb$_{2}$O$_{6}$ in which there are no strong electron correlations.\cite{yonezawa04} Deviation from standard Bloch-Gr\"{u}neisen theory is often observed in low-dimensional metals.\cite{oshiyama78} This explanation is, however, not applicable in our cases because KSbO$_{3}$-type rhenates are isotropic electron systems with a cubic symmetry. Scattering by optic phonons with low excitation energy might be responsible for the anomalous temperature dependence of the resistivities. The characteristic cage-like structure of the KSbO$_{3}$-type rhenates would have a vibration mode with low exicitation energy. 

The resistivity of Bi$_{3}$Re$_{3}$O$_{11}$ exhibits another type of anomalous temperature dependence. In high temperature region, there is no large deviation from the behavior of ordinary metals. However, the $\rho$ vs $T$ curve shows a round bend at 50 K, below which the slope becomes steeper again. Remind that the susceptibility of Bi$_{3}$Re$_{3}$O$_{11}$ after the subtraction of Curie-Weiss function shows anomalous temperature dependence below 50 K. This might indicate that the DOS has a pseudo-gap like dip at $E_{F}$. In that case, the slope of the $\rho$ vs $T$ curve becomes smaller above 50 K because additional carriers are created by thermal excitations. Although their origins have not been clarified, we should note that similar gap-like behaviors in magnetic susceptibility appear in several KSbO$_{3}$-type ruthenates, La$_{4}$Ru$_{6}$O$_{19}$\cite{khalifah01} and (Ba$_{1-x}$Sr$_{x}$)$_{2}$Ru$_{3}$O$_{9}$\cite{sato04}. Khalifah suggested that the singular band structure based on the molecular orbitals of Ru-Ru dimer might be responsible of the gap-like magnetism in La$_{4}$Ru$_{6}$O$_{19}$.\cite{khalifah01b} In our case, it seems to be strange that only Bi$_{3}$Re$_{3}$O$_{11}$ has anomalous feature because there are no large difference in the Re-Re interatomic distance among our rhenates. Band calculations and more experiments, such as specific-heat measurements, will clarify this problem.

\section{Conclusion}
We synthesized single crystals of several KSbO$_{3}$-type rhenate by a hydrothermal method and investigated their magnetism and electric conduction. The results shows metallic electronic states. Resistivities of Sr$_{2}$Re$_{3}$O$_{9}$, Pb$_{6}$Re$_{6}$O$_{19}$ and La$_{4}$Re$_{6}$O$_{19}$ has anomalous  $\rho=\rho_{0}+AT^{n}$ $(n \approx 1.6)$ temperature dependence. The magnetic susceptibilities exhibit a Pauli paramagnetism with a Curie-Weiss contribution, probably from impurities. Bi$_{3}$Re$_{3}$O$_{11}$ also exhibits metallic behavior, but the resistivity and the magnetic susceptibility after the subtraction of the Cuire-Weiss component has anomaly around 50 K.

\section*{Acknowledgment}
We would like to thank Dr. K. Miyagawa and Prof. K. Kanoda for SQUID measurements. Figures \ref{fig2} and \ref{fig3} was drawn with VENUS developed by Dilanian and Izumi. This study was supported by Grants-in-Aid for Scientific Research No. 15750127 and No. 17540341 from the Ministry of Education, Culture, Sports, Science and Technology.


\begin{thebibliography}{99} 
\bibitem{ferretti65} A. Ferretti, D. B. Rogers and J. B. Goodenough: J. Phys. Chem. Solids \textbf{26} (1965) 2007.
\bibitem{tanaka76} T. Tanaka, T. Akahane, E. Bannai, S. Kawai, N. Tsuda and Y. Ishizawa: J. Phys. C \textbf{9} (1976) 1235.
\bibitem{mattheiss69} L. F. Mattheiss: Phys. Rev. \textbf{181} (1969) 987.

\bibitem{sakai01} H. Sakai, K. Yoshimura, H. Ohno, H. Kato, S. Kambe, R. E. Walstedt, T. D. Matsuda and Y. Haga: J. Phys: Cond. Mat. \textbf{13} (2001) L785.
\bibitem{hanawa01} M. Hanawa, Y. Muraoka, T. Tayama, T. Sakakibara, J. Yamaura, Z. Hiroi: Phys. Rev. Lett. \textbf{87} (2001) 187001.

\bibitem{chen02} S. Chen and H. B\"{u}ttner: Eur. Phys. J. B \textbf{29} (2002) 15.
\bibitem{shastry81} B. S. Shastry and B. Sutherland: Physica \textbf{108B} (1981) 1069.

\bibitem{kageyama99} H. Kageyama, K. Yoshimura, R. Stern, N. V. Mushnikov, K. Onizuka, M. Kato, K. Kosuge,
 C.P. Slichter, T. Goto, and Y. Ueda: Phys. Rev. Lett. \textbf{82} (1999) 3168.
\bibitem{miyahara03} S. Miyahara and K. Ueda: J. Phys.: Condens. Matter. \textbf{15} (2003) R327.

\bibitem{khalifah01} P. Khalifah, K. D. Nelson, R. Jin, Z. Q. Mao, Y. Liu, Q. Huang, X. P. A. Gao, A. P. Ramirez,
 and R. J. Cava: Nature \textbf{411} (2001) 669.
\bibitem{khalifah01b} P. Khalifah and R. J. Cava: Phys. Rev. B \textbf{64} (2001) 085111.
\bibitem{fujita03} T. Fujita, K. Tsuchida, Y. Yasui, Y. Kobayashi and M. Sato: Physica B \textbf{329-333} (2003) 743. 
\bibitem{tsuchida04} K. Tsuchida, C. Kato, T. Fujita, Y. Kobayashi and M. Sato: J. Phys. Soc. Jpn. \textbf{73} (2004) 698.

\bibitem{sato04} H. Sato, T. Watanabe and J. -I. Yamaura: Solid State Commun. \textbf{131} (2004) 707.

\bibitem{baud79} G. Baud, J. P. Besse, R. Chevalier and B. L. Chamberland: J. Solid State Chem. \textbf{28} (1979) 157.
\bibitem{abakumov98} A. M. Abakumov, R. V. Shpanchenko and E. V. Antipov: Z. Anorg. Allg. Chem. \textbf{624} (1998) 750.
\bibitem{longo68} J. M. Longo and A. W. Sleight: Inorg. Chem. \textbf{7} (1968) 108.
\bibitem{morrow68} N. L. Morrow and L. Katz: Acta Cryst. B \textbf{24} (1968) 1466.
\bibitem{bramnik98} K.G. Bramnik, A. M. Abakumov, R. V. Shpanchenko, E. V. Antipov and G. Van Tendeloo: J. Alloys Cmpds. \textbf{278} (1998) 98.
\bibitem{jeitschko99} W. Jeitschko, D. H. Heumannsk\"{a}mper, M. S. Schriewer-P\"{o}ttgen and U. Ch. Rodewald: J. Solid State Chem. \textbf{147} (1999) 218.
\bibitem{cheetham81} A. K. Cheetham and A. R. Rae-Smith: Mater. Res. Bull. \textbf{16} (1981) 7.

\bibitem{sasaki06} A. Sasaki, M. Wakeshima and Y. Hinatsu: J. Phys.: Condens. Matter \textbf{18} (2006) 9031.

\bibitem{kawamura04} Y. Kawamura and H. Sato: J. Alloys Compds. \textbf{383} (2004) 209.
\bibitem{yonezawa04} S. Yonezawa and Y. Maeno: Phys. Rev. B \textbf{70} (2004) 184523.

\bibitem{oshiyama78} A. Oshiyama, K. Nakao and H. Kamimura: J. Phys. Soc. Jpn. \textbf{45} (1978) 1136.

\end{thebibliography}
\end{document}